\documentclass[aps,prl,twocolumn,nofootinbib]{revtex4-1}

\usepackage{url,comment}
\usepackage{times}
\usepackage{latexsym}
\usepackage{graphicx, graphics, hyperref, amsmath, amssymb, slashed, xcolor, bbm,amsthm, array}
 \usepackage{subfigure}

\usepackage{rotating}
\usepackage{afterpage}

\newcommand{\nc}{\newcommand}
\nc{\beq}{\begin{equation}}
\nc{\eeq}{\end{equation}}
\nc{\barray}{\begin{eqnarray}}
\nc{\earray}{\end{eqnarray}}
\nc{\barrayn}{\begin{eqnarray*}}
\nc{\earrayn}{\end{eqnarray*}}
\nc{\bcenter}{\begin{center}}
\nc{\ecenter}{\end{center}}
\nc{\mc}{\mathcal}
\nc{\er}[1]{(\ref{eq:#1})}
\nc{\onehalf}{\frac{1}{2}} 
\nc{\partialbar}{\bar{\partial}}
\nc{\psit}{\widetilde{\psi}}
\nc{\Tr}{\mbox{Tr}}
\nc{\hc}{\mbox{H.c.}}
\nc{\ev}{\;\mathrm{eV}}
\nc{\mev}{\;\mathrm{MeV}}
\nc{\gev}{\;\mathrm{GeV}}
\nc{\tev}{\;\mathrm{TeV}}

\def\chii0{\chi_i^0}
\def\chij0{\chi_j^0}

\newcommand{\gsim}{\lower.7ex\hbox{$\;\stackrel{\textstyle>}{\sim}\;$}}
\newcommand{\lsim}{\lower.7ex\hbox{$\;\stackrel{\textstyle<}{\sim}\;$}}
\nc{\ttbar}{t\bar t}
\def\ifb{{\ \rm fb}^{-1}}

\newcommand{\fref}[1]{Fig.~\ref{f.#1}}
\newcommand{\eref}[1]{Eq.~(\ref{e.#1})}

\newcommand{\cref}[1]{Chapter~\ref{c.#1}}

\graphicspath{{plots/}}

\begin{document}

\title{
Quirky Explanations for the Diphoton Excess
}

\author{David Curtin}
\email{dcurtin1@umd.edu}
\author{Christopher B. Verhaaren}
\email{cver@umd.edu}

\affiliation{Maryland Center for Fundamental Physics, Department of Physics,\\ University of Maryland, College Park, MD 20742-4111 USA}

\date{\today}
\begin{abstract}
We propose two simple quirk models to explain the recently reported 750 GeV diphoton excesses at ATLAS and CMS. It is already well-known that a real singlet scalar $\phi$ with Yukawa couplings $\phi  \bar X X$ to vector-like fermions $X$ with mass $m_X > m_\phi/2$ can easily explain the observed signal, provided $X$ carries both SM color and electric charge. We instead consider first the possibility that the pair production of a fermion, charged under both SM gauge groups and a confining $SU(3)_v$ gauge group, is responsible. If pair produced it forms a quirky bound state, which promptly annihilates into gluons, photons, v-gluons and possibly SM fermions. This is an extremely minimal model to explain the excess, but is already in some tension with existing displaced searches, as well as dilepton and dijet resonance bounds. We therefore propose a hybrid Quirk-Scalar model, in which the fermion of the simple $\phi \bar X X$ toy model is charged under the additional $SU(3)_v$ confining gauge group. Constraints on the new heavy fermion $X$ are then significantly relaxed. The main additional signals of this model are possible dilepton, dijet and diphoton resonances at $\sim 2 \tev$ or more from quirk annihilation, and the production of v-glueballs through quirk annihilation and $\phi$ decay. The glueballs can give rise to spectacular signatures, including displaced vertices and events with leptons, photons and $Z$-bosons. If the Quirk-Scalar model is responsible for the 750 GeV excess it should be discovered in one of these channels with 20 or $300\ifb$ of LHC run 2 data.
\end{abstract}



\maketitle

Recently, both of the major LHC collaborations announced excesses in events containing two photons, near a diphoton invariant mass of around 750 GeV. The ATLAS search \cite{atlas750} used 3.2 $\ifb$ of data at $\sqrt{s} = 13 \tev$ and reports a local (global) excess significance of $3.9\sigma$  ($2.3 \sigma$). The CMS search \cite{CMS:2015dxe} used 2.6 $\ifb$ and reports a local excess significance of $2.6\sigma$.

This excess has already generated considerable interest in the theory community \cite{DiChiara:2015vdm,
Franceschini:2015kwy, 
Pilaftsis:2015ycr, 
Buttazzo:2015txu, 
Harigaya:2015ezk, 
Knapen:2015dap, 
Angelescu:2015uiz,
McDermott:2015sck,
Ellis:2015oso,
Gupta:2015zzs,
Backovic:2015fnp,
Mambrini:2015wyu,
Knapen:2015dap,
Nakai:2015ptz,
Higaki:2015jag,
Low:2015qep,
Bellazzini:2015nxw,
Petersson:2015mkr,
Molinaro:2015cwg}\footnote{
Refs. \cite{Cao:2015pto, Matsuzaki:2015che, Kobakhidze:2015ldh, Cox:2015ckc, Ahmed:2015uqt, Agrawal:2015dbf, Martinez:2015kmn, Becirevic:2015fmu, No:2015bsn, Demidov:2015zqn, Chao:2015ttq, Fichet:2015vvy, Bian:2015kjt, Chakrabortty:2015hff, Csaki:2015vek,  Falkowski:2015swt, Aloni:2015mxa, Bai:2015nbs} on this subject appeared at the same time as this letter.}. If a Beyond Standard Model (BSM) process is responsible, compatibility with LHC run 1 searches favors gluon or heavy quark initiated production. The best-fit width is around 45 GeV but a very narrow resonance is also compatible with the data. Events in the signal bins do not appear kinematically very different from the sidebands, somewhat disfavoring explanations that involve the production of additional missing transverse energy (MET), leptons, etc. This motivates interpreting the signal in the context of a minimal benchmark model, where a Standard Model (SM) singlet scalar $\phi$ with mass around 750 GeV is produced through gluon-fusion and decays to two photons $g g \to \phi \to \gamma \gamma$. 
Taking acceptance into account, the approximate cross section corresponding to the excess \cite{Franceschini:2015kwy} is $\sigma(p p \to \phi \to \gamma \gamma) \approx (6 \pm 3) \mathrm{fb}$ at CMS and $(10 \pm 3) \mathrm{fb}$ at ATLAS with $\sqrt{s} = 13 \tev$.

This excess is intriguing for several reasons. A SM-neutral scalar couples to gluons (photons) via a dimension-5 effective operator, which can be generated by loops of colored (electrically charged) particles with sizable couplings to the scalar. In order for the diphoton decay to be observable, the scalar cannot have any large tree-level couplings to other SM particles. This excludes gluon fusion via a SM top loop as the production mechanism, since otherwise $\phi \to t \bar t$ would completely dominate. The diphoton excess, if its BSM origin is confirmed, might therefore imply the existence of additional matter states with color and electric charge. 

A simple toy model can be constructed by adding the 750 GeV singlet scalar $\phi$ and a new vector-like fermion $X$  to the SM. $X$ is an $SU(3)_c$ fundamental, carries electric charge $Q_X$, and is heavier than about 380 GeV to avoid tree-level decays of $\phi$.  The simple interaction Lagrangian
\begin{equation}
\label{e.phiXX}
\mathcal{L_\mathrm{int}} = y_X \phi \bar X X \ ,
\end{equation}
where $\phi$ does not couple directly to any SM fields, can then give rise to the observed diphoton signal for $m_X \sim \tev$, $y_X \sim 1$ and $Q_X \sim 1$. Several varieties of this model have already been explored 
\cite{DiChiara:2015vdm,
Franceschini:2015kwy, 
Pilaftsis:2015ycr, 
Buttazzo:2015txu, 
Harigaya:2015ezk, 
Knapen:2015dap, 
Angelescu:2015uiz,
McDermott:2015sck,
Ellis:2015oso,
Gupta:2015zzs}, even before 13 TeV LHC data was available \cite{Jaeckel:2012yz}.

The existence of new colored and electrically charged matter states around a TeV has many important consequences, which must be considered in the context of this diphoton excess. If $X$ lives long enough to escape the detector it has to be heavier than about  $\sim 1.1 \tev$ \cite{CMS:2015kdx}. A stable TeV-scale colored fermion would have serious implications for cosmology, and making it unstable adds considerable complication for the model, which would still be subject to direct search constraints.

In this letter, we examine two very simple models to explain the diphoton excess. In both cases, we add a vector-like fermion $X$ and a new confining gauge group $SU(3)_v$ to the Standard Model. $X$ is a fundamental under both $SU(3)_c$ and $SU(3)_v$ and carries hypercharge:
\begin{equation}
\begin{array}{|cc|c|c|c|c|}
\hline
& & SU(3)_v & SU(3)_c & SU(2)_L & U(1)_Y\\
\hline
X & & \mathbf{3} & \mathbf{3} & \mathbf{1} & Q_X\\
\hline
\end{array}
\end{equation}
Assuming no lighter states with $SU(3)_v$ charge, this new gauge group confines in the IR and realizes a pure-glue Hidden Valley \cite{Strassler:2006im, Strassler:2006ri, Strassler:2006qa, Han:2007ae}, with v-glueballs making up the low-energy hidden hadron spectrum \cite{Morningstar:1999rf}. (We assume the confinement scale is significantly below $m_X$.) When $X$ is pair produced via its SM color charge it forms a \emph{quirky bound state} \cite{Okun:1980kw,Okun:1980mu,Kang:2008ea}, since the v-gluon string connecting them cannot break by exciting light quark pairs out of the vacuum. For $SU(3)_v$ confinement scale $\Lambda_v \gtrsim \Lambda_\mathrm{QCD}$, the bound state promptly de-excites via emission of soft quanta like photons and gluons \cite{Kang:2008ea,Burdman:2008ek,Cheung:2008ke,Harnik:2008ax,Harnik:2011mv,Fok:2011yc}, and annihilates into v-gluons, SM gluons and photons, and possibly SM fermions. The v-gluons hadronize to form v-glueballs, which decay mostly to SM gauge bosons via dimension-8 operators that are generated by loops of the bifundamental $X$ \cite{Juknevich:2009ji, Juknevich:2009gg}. This gives rise to a variety of signatures at colliders, including displaced vertices.

The first model we examine adds only the fermion $X$ to the SM.\footnote{A similar possibility was discussed in ref.~\cite{Agrawal:2015dbf}, which appeared concurrently with this letter.} It is pair produced with strong cross section, and the quirk annihilation can produce the observed diphoton excess if $m_X \approx 370 \gev$. This scenario is intriguing and minimal. However, as we will show, the feasibility of this model is at best marginal for a variety of reasons.

The second model we examine involves modifying the toy model of \eref{phiXX} and making $X$ a fundamental under $SU(3)_v$. The quirky nature of $X$ makes it cosmologically safe and avoids various bounds, including long-lived charged particle searches. We show that this model can explain the excess, and is compatible with current bounds while being potentially discoverable via additional heavy resonances and its glueball signatures at the LHC run 2.

This paper is structured as follows. Before discussing the models in detail we review the phenomenology of quirks and glueballs. We then examine the pure quirk model, describe how it can accommodate the diphoton excess and how it is constrained by other searches. Since this model is only marginally feasible we then introduce the scalar-quirk hybrid model, and discuss its diphoton signal and other discovery channels at the LHC run 2.

\textbf{QUIRK PHENOMENOLOGY --- } The vector-like quirks $X$ can be pair-produced at the LHC through their color charge, with a cross section that is shown in \fref{xsec}.

\begin{figure}
\begin{center}
\includegraphics[width=0.45\textwidth]{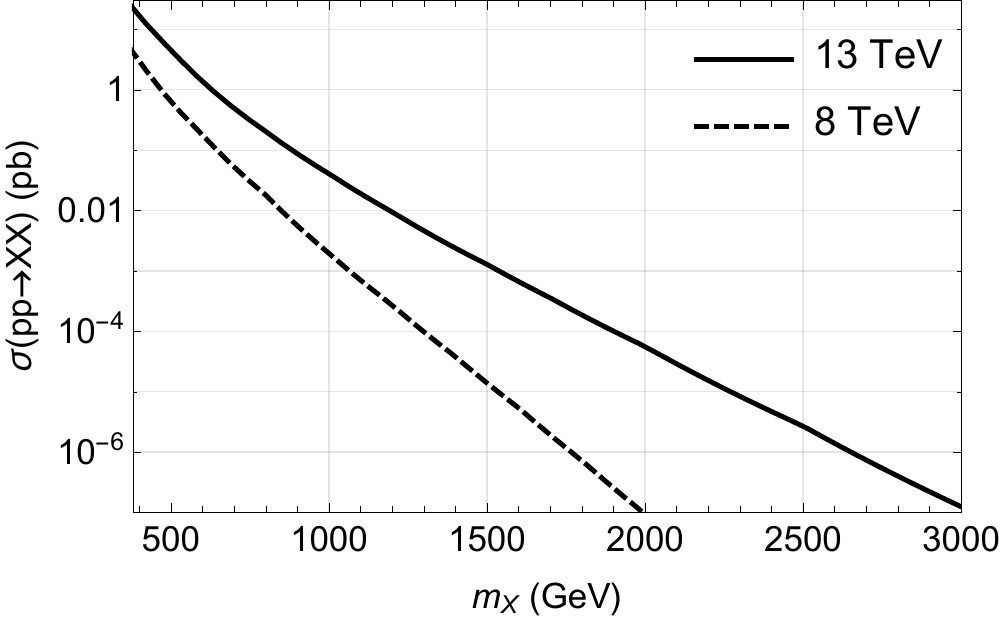}
\end{center}
\caption{
$\bar X X$ pair production cross section at the 8 and 13 TeV LHC. Computed at LO in MadGraph \cite{Alwall:2014hca}.
}
\label{f.xsec}
\end{figure}

When the $\bar X X$ pair is created it forms an excited quirky bound state \cite{Kang:2008ea,Burdman:2008ek,Cheung:2008ke,Harnik:2008ax,Harnik:2011mv,Fok:2011yc}. It promptly de-excites by emitting soft quanta like gluons and photons, which randomize its orbital angular momentum. Since $s$-wave annihilation is very strongly preferred, this suppresses annihilation until the quirk has de-excited to near its lowest lying states, which is a distribution of quirkonium bound states in analogy to SM quarkonia. 

The relevant quirkonium annihilation widths can be adapted from \cite{Barger:1987xg, Cheung:2008ke}. Before annihilation, the bound state can be either in a spin singlet or triplet state. The decay width of the triplet is only $\sim 3\%$ of the singlet. Therefore, if soft emissions randomize the quirk between the singlet and triplet state while it de-excites and annihilates, the singlet decay branching fractions will dominantly deteremine the final states. If, on the other hand, the singlet and triplet states are democratically populated after soft emission ceases, but well before annihilation occurs, then both singlet and triplet decay branching fractions determine the final state in comparable proportion. The large uncertainties in understanding this strongly coupled bound state make it difficult to make a definitive determination which possibility is realized. We will discuss both cases.

Assume first that the spin singlet state ${}^1S_0$ dominates annihilation. The branching fractions are very insensitive to mass. For $SU(3)_v$ coupling $\alpha_v \sim \alpha_s$ (evaluated at $\mu = 2 m_X$), $m_X \sim \tev$,  and $Q_X \lesssim 2$, they are 
\begin{eqnarray}
\nonumber 
\mathrm{Br}({}^1S_0 \to \gamma \gamma) &\approx& 7 \ \cdot \ 10^{-3} \ \cdot \ Q_X^4
\\
\nonumber
\mathrm{Br}({}^1S_0 \to Z \gamma) &\approx& 4 \ \cdot \ 10^{-3} \ \cdot \ Q_X^4
\\
\mathrm{Br}({}^1S_0 \to Z Z) &\approx& 6 \ \cdot \ 10^{-4} \ \cdot \ Q_X^4
\\
\nonumber 
\mathrm{Br}({}^1S_0 \to gg) &\approx& \frac{1}{1+(\alpha_v/\alpha_s)^2}
\\
\nonumber 
\mathrm{Br}({}^1S_0 \to g'g') &\approx& \frac{(\alpha_v/\alpha_s)^2}{1+(\alpha_v/\alpha_s)^2}  \ ,
\end{eqnarray}
where $g'$ are v-gluons. The $\alpha_v/\alpha_s$ ratio can be expressed in terms of the lightest glueball mass $m_0 \approx 7 \Lambda_v$ assuming pure gauge RG evolution between $m_0$ and $2 m_X$. As shown in \fref{alphavoveralphas}, this favors ratios in the range $\sim$ 0.7 - 2.5. Note that there is no gluon-mediated $\bar X X \to \bar q q$ decay, since the quirks form a color singlet state.

The diphoton final state could generate the diphoton excess observed at ATLAS and CMS, as discussed below. The gluon final state results in a dijet resonance. 
The potentially most exciting additional signal of quirk production is their annihilation into v-gluons, which hadronize into jets of glueballs.  A similar signature has been recently discussed in the context of Neutral Naturalness \cite{Chacko:2015fbc}. Details of hadronization in a pure $SU(3)$ gauge theory are currently unknown, which makes detailed study of this signal challenging. However, the glueballs produced in quirk annihilation may give rise to displaced vertices, lepton pairs, resonant and non-resonant photons, and dijets. We study some possible aspects of glueball phenomenology below. Even without detailed knowledge of hadronization, several important predictions can be made.

If both the spin singlet and triplet ${}^3S_1$ democratically determine the branching fraction, the situation is quite different. The triplet state can mix with the SM $\gamma/Z$ and decay to SM fermions $\bar f f$. For $m_X \sim \tev$, the branching fractions are very well described by
\begin{eqnarray}
\nonumber 
\sum_i \mathrm{Br}({}^3S_1 \to \bar f_i f_i) &\approx&  \frac{Q_X^2}{Q_X^2 \ + \  0.05 \  \left(1 + (\alpha_v/\alpha)^3\right)}
\\
\mathrm{Br}({}^3S_1 \to  g g g) &\approx&  \frac{0.05}{Q_X^2 \ + \  0.05 \  \left(1 + (\alpha_v/\alpha)^3\right)}
\\
\nonumber 
\mathrm{Br}({}^3S_1 \to  g' g' g' ) &\approx&  \frac{0.05 \ (\alpha_v/\alpha)^3}{Q_X^2 \ + \  0.05 \  \left(1 + (\alpha_v/\alpha)^3\right)} \ .
\end{eqnarray}
(Here we have omitted for simplicity  the strongly subdominant $Zh$, $Zg^{(\prime)}g^{(\prime)}$, and $\gamma g^{(\prime)}g^{(\prime)}$ decay modes.)
The branching fraction to SM fermions is $\sim 50\%$ for $Q_X \sim 1/3$ and $\alpha_v/\alpha\sim$ 1 and quickly becomes dominant as $Q_X$ is increased. As determined by hypercharges, 27\% of the produced SM fermions are $e^+ e^-$ or $\mu^+ \mu^-$. Therefore, dilepton resonance searches could strongly constrain light quirks if a significant fraction annihilates in the triplet state.

Given the uncertainties, the total width of the quirk annihilation signature cannot be exactly determined. However, for $m_X \sim $ TeV and $\alpha_v \sim \alpha_s$, the individual quirkonium states have $\sim$ MeV widths or below,  with $\sim$ GeV splittings between the states \cite{Kang:2008ea, Fok:2011yc}. Annihilation from these states would therefore lead to at most few-GeV-scale resonance widths, but this could be broadened if annihilation proceeds while the quirks are still somewhat excited.

\begin{figure}
\begin{center}
\includegraphics[width=0.45\textwidth]{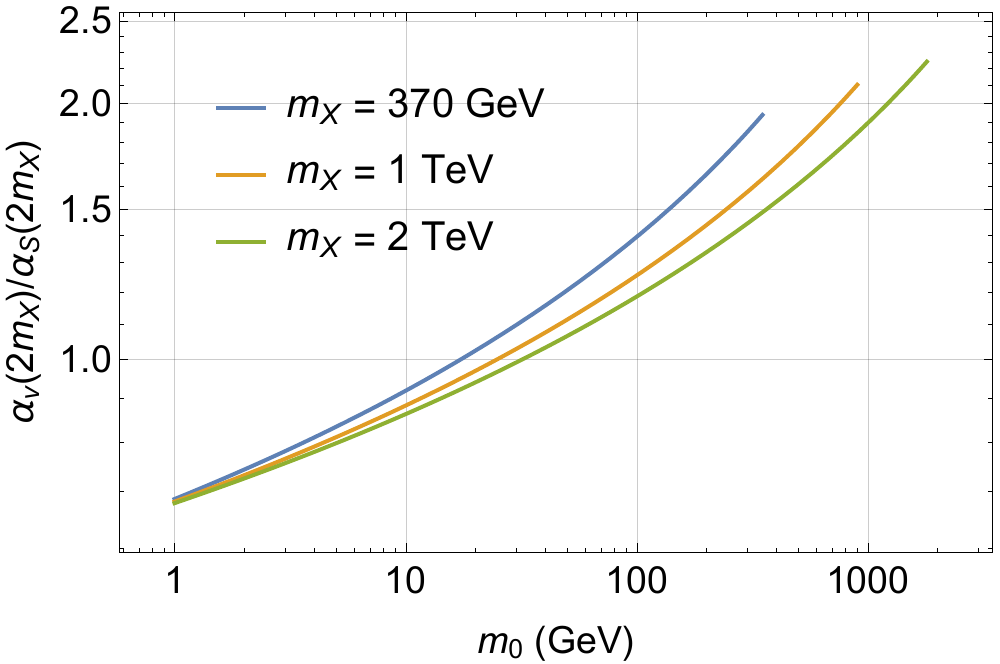}
\end{center}
\caption{
Ratio of $\alpha_v/\alpha_s$ evaluated at $\mu = 2 m_X$ as a function of glueball mass $m_0 \approx 7 \Lambda_v$ derived using the pure $SU(3)_v$ gauge RGE, and assuming $m_0 < m_X$.
}
\label{f.alphavoveralphas}
\end{figure}

\textbf{GLUEBALL PHENOMENOLOGY --- } A pure $SU(3)_v$ confining gauge theory has $\sim 12$ stable glueball states \cite{Morningstar:1999rf}. The lightest state is the singlet $0^{++}$ with a mass of $m_0 \approx 7 \Lambda_v$, while the heaviest state has a mass of $\approx 2.8 m_0$. The $X$ acts as a bifundamental messenger, being charged under both SM gauge groups and $SU(3)_v$. Loops of $X$ allow the glueballs to decay to SM gauge bosons and in some cases fermion pairs via dimension-8 operators, with decay rates that have been computed by \cite{Juknevich:2009ji, Juknevich:2009gg} in terms of hadronic matrix elements that can be extracted from lattice calculations. (Note that $X$ carries no $SU(2)_L$ charge.) The glueball lifetime can be very long and  is subject to significant uncertainties due to several uncomputed hadronic matrix elements. Furthermore, there are potentially many orders of magnitude of lifetime difference between the different glueball states. Finally, the details of hadronization of v-gluons are unknown. While it is possible to parameterize our ignorance in this regard \cite{Chacko:2015fbc}, here we merely point out that a significant fraction of the produced glueballs is expected to be in the lightest $0^{++}$ state \cite{JuknevichPhD}, and that the lifetime of this state is relatively well known.

\begin{figure}
\begin{center}
\includegraphics[width=0.45\textwidth]{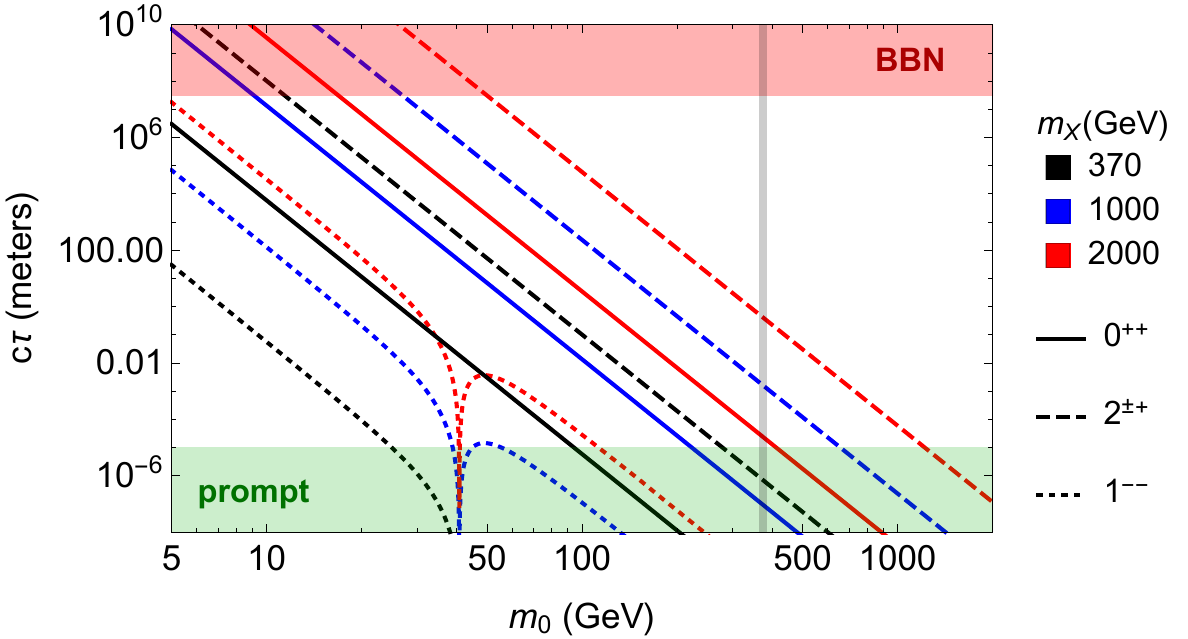}
\end{center}
\caption{
The range of possible glueball decay lengths as a function of $m_0$, the mass of the lightest $0^{++}$ state. Red band: lifetimes longer than 0.1 seconds may conflict with BBN bounds. Green band: prompt decay at colliders with $c \tau < 10 \mu\mathrm{m}$. Vertical gray line: above this mass, a 750 GeV scalar cannot decay into glueballs. Solid lines: decay length of the $0^{++}$ state. Dashed (dotted) lines: upper (lower) estimate for $2^{\pm+}$ ($1^{--}$) glueballs. The other glueball lifetimes should lie roughly between the dashed and dotted lines. Evaluated for $m_X = 370 \gev$ (black), 1 TeV (blue) and 2 TeV (red).
}
\label{f.glueball}
\end{figure}

\fref{glueball} shows the range of possible glueball lifetimes as a function of $m_0$, the mass of the lightest $0^{++}$ state, for $m_X = 370 \gev$ (black), $1 \tev$ (blue) and 2 TeV (orange). The green shaded region indicates decay that is prompt on collider scales, $c \tau < 10 \mu\mathrm{m}$. The red shaded region indicates lifetimes longer than 0.1 second, which is problematic for Big Bang Nucleosynthesis. The vertical gray band indicates $m_0 = 375 \gev$, above which a 750 GeV scalar cannot decay to v-gluons.

The solid lines indicate the reasonably well-known lifetime of the $0^{++}$ glueball, which should be commonly produced in v-hadronization. 
The other glueball lifetimes are contained between the dashed and dotted lines. In cases where the hadronic matrix elements have not been computed on the lattice we employ the estimates of \cite{Juknevich:2009gg} varied within a factor of $1/3 - 3$. 
The dashed lines indicate the lifetime of the longest-lived $2^{++}$ or $2^{-+}$ state with lower estimates for the unknown hadronic matrix elements. The dotted lines indicate the lifetime of the shortest-lived $1^{--}$ state with upper estimates for the unknown hadronic matrix elements. (The resonance feature at $m_0 \approx 40 \gev$ occurs when $m_{1^{--}} \approx 2.2 m_0 \approx m_Z$.) Most of the glueballs decay radiatively to other glueballs via emission of photons, or directly to SM gauge bosons. Decays to gluons dominate, with $\sim 10^{-4} - 10^{-2}$ branching fractions to $\gamma \gamma$, $\gamma Z$ and $Z Z$ (when kinematically accessible). An interesting exception is the $1^{--}$ state, which can mix with the SM $Z$-boson and have significant branching fraction into SM fermion pairs, including leptons.

This basic discussion of glueball phenomenology allows us to evaluate the feasibility of our quirky toy models for the diphoton excess.

\textbf{PURE QUIRK MODEL ---} The pure quirk model, as outlined above, is defined by merely adding $X$ and $SU(3)_v$ to the SM. In order to produce the diphoton excess, the mass of $X$ has to be about 370 GeV, which we assume here. At that mass, the 13 TeV (8 TeV) LHC production cross section is about 26 pb (5pb) at lowest order, which is sufficient for this discussion. Coincidentally, this leads to about the same number of $\bar X X$ pairs produced at 13 TeV with $3.2 \ifb$ as at 8 TeV with $20\ifb$, which simplifies our discussion. 

We first assume that the quirk annihilates in the spin singlet state. In that case, \fref{quirkNgaga} shows the number of $\gamma \gamma$, $Z \gamma$ and $ZZ$ events produced through quirk pair production and annihilation at the 13 TeV LHC with $3.2 \ifb$. 
The ATLAS excess favors $\sim 20 - 40$ events, which is easily accommodated for $Q_X \sim 1/3 - 1/2$. However, this model produces a similar number of diphoton events at run 1, placing it in tension with those searches. 

In this range of $Q_X$, quirks produce only a few $ZZ$ events, which is safe from run 1 constraints like \cite{Khachatryan:2014gha} but may be discovered in future searches. The number of $Z \gamma$ events is smaller but comparable to $\gamma \gamma$. This is well within 8 TeV limits \cite{Aad:2014fha} but potentially detectable with more run 2 data.

Apart from the large diphoton rate at run 1, the most important constraints on this model with pure spin-singlet annihilation actually stem from the gluon and v-gluon final states of quirk annihilation. 

Annihilation into gluons occurs some $\mathcal{O}(1)$ fraction of the time, and can be constrained by dijet resonance searches. These searches are challenging due to the large event rates, but CMS performed an un-prescaled dijet resonance search using data scouting techniques \cite{CMS:2015neg} at run 1. At a dijet mass of 750 GeV, this search gives the constraint
\begin{equation}
\sigma(p p \to \bar X X) \ \cdot \ \mathrm{Br}(\bar X X \to g g) \ \lesssim 1 \ \mathrm{pb}
\end{equation}
Since the 8 TeV production cross section is about 5 pb, this already constraints $\mathrm{Br}(\bar X X \to g g)  \lesssim 0.2$, which implies $\alpha_v/\alpha_s \gtrsim 2$. Such a large ratio for $m_X =  370 \gev$ can only be achieved if the glueballs are almost as heavy as the $X$ fermions themselves, see \fref{alphavoveralphas}.

\begin{figure}
\begin{center}
\includegraphics[width=0.4\textwidth]{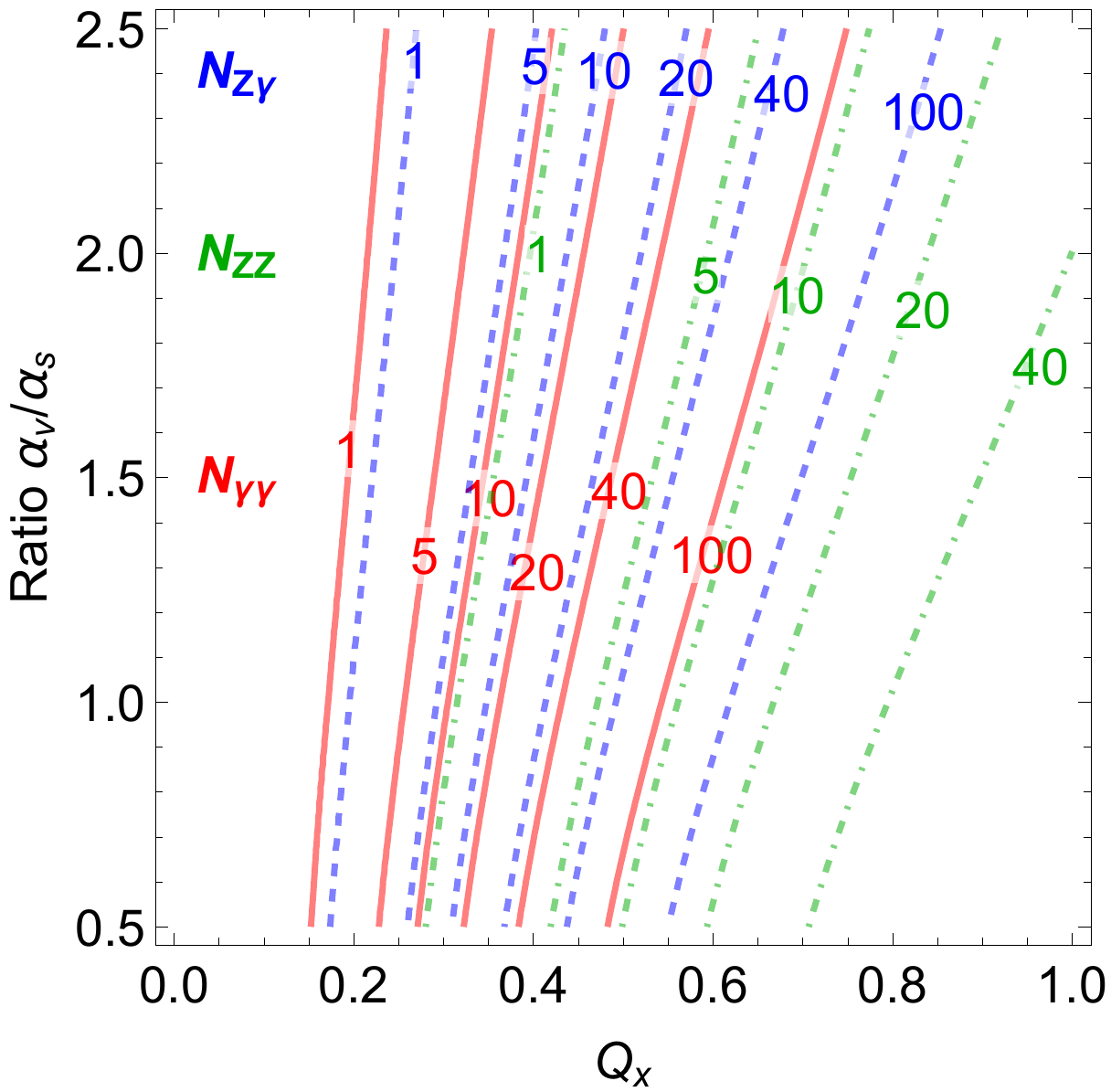} 
\end{center}
\caption{
Number of $\gamma \gamma$ (red solid), $Z \gamma$ (blue dashed) and $ZZ$ (green dot-dashed) events produced at the 13 TeV LHC with $3.2 \ifb$ from $\bar X X$ pair production and quirky annihilation. Shown in the plane of $X$ electric charge $Q_X$ and the ratio $\alpha_v/\alpha_s$ evaluated at $\mu = 2m_X$, which can be mapped to the mass of the lightest glueball $m_0$, see \fref{alphavoveralphas}.
}
\label{f.quirkNgaga}
\end{figure}

Annihilation into v-gluons yields jets of v-glueballs, which can decay in the detector with measurable displacement. This allows us to estimate constraints using the CMS displaced dijet search at run 1 \cite{CMS:2014wda}. In the $m_0 \sim \mathcal{O}(100 \gev)$ range favored by the dijet bounds, the $0^{++}$ glueballs decay promptly into gluons. Since they are presumably a sizable fraction of the produced glueballs, and given that the $\bar X X$ state has an invariant mass of around 750 GeV, the resulting  prompt jets should surpass the 
$H_T > 325 \gev$ threshold of this search and lead to large signal acceptance when any other glueball decays in the tracker. In this range of $m_0$, the $2^{\pm+}$ glueball state has lifetimes in the cm range, see \fref{glueball}. This means that events with $2^{\pm+}$ glueballs should be detected by this analysis.  For cm-scale lifetimes, the limit can be translated as roughly 
\begin{equation}
\sigma(p p \to \bar X X) \ \cdot \ \mathrm{Br}(\bar X X \to 2^{\pm+}0^{++} +  \ldots) \lesssim 10^{-3} \ \mathrm{pb}
\end{equation}
at 8 TeV. That means that less than $\sim$  0.1\% of $\bar X X$ pair production events can contain any glueballs which decay on cm scales like the $2^{\pm +}$. While no details on v-gluon hadronization are known, this nevertheless appears to be a fairly stringent requirement.

When a fraction $R_{3/1}$  of quirks annihilate in the spin triplet state, the diphoton rate described above is correspondingly reduced. Even so, the pure quirk scenario could still describe the diphoton excess quite easily with slightly increased $Q_X$. The far more significant change is the dilepton resonance signature of quirk production. For $m_X = 370 \gev$ and $Q_X = 1/2$, the triplet branching fraction to electron and muon pairs is about 20\% (8\%) for $\alpha_v/\alpha = 0.7$ (2).

The sensitivity of dilepton resonance searches at 750 GeV from LHC run 1 searches \cite{Aad:2014cka, Khachatryan:2014fba} is around $10^{-3}$ pb. This translates to a bound on the fraction of quirks that annihilate in the triplet state:
\begin{equation}
R_{3/1} \lesssim 10^{-3} \ .
\end{equation}
(The bound derived from the run 2 searches with $\sim 3 \ifb$ \cite{atlas8tevdilepton, CMS:2015nhc} is nearly identical.) This seems like another very stringent requirement for the pure quirk model to satisfy. 

Given the tension between the quirk-produced diphoton signal at 13 TeV and the lack of a large excess at 8 TeV, as well as the lack of signals in the CMS displaced dijet analysis and dilepton channels, we conclude that this model is at best marginally viable as an explanation for the diphoton excess. However, if direct quirk pair production is responsible, it should show up with 20 $\ifb$ of LHC run 2 data as  a dijet or dilepton resonance peak, or a displaced signal.

\textbf{QUIRK-SCALAR MODEL --- }
We now modify the original toy model of \eref{phiXX} by charging the vector-like fermion X under the $SU(3)_v$ gauge group. The BSM particle content is therefore 
\begin{equation}
\begin{array}{|cc|c|c|c|c|}
\hline
& & SU(3)_v & SU(3)_c & SU(2)_L & U(1)_Y\\
\hline
X & & \mathbf{3} & \mathbf{3} & \mathbf{1} & Q_X\\
\hline
\phi & & \mathbf{1} & \mathbf{1} & \mathbf{1} &  0 \\
\hline
\end{array}
\end{equation}
$\phi$ is taken to be a real singlet scalar. The minimal additional non-kinetic terms of the Lagrangian are
\begin{equation}
\label{e.L}
\mathcal{L} = \frac{1}{2} m_\phi^2 \phi^2 + m_X \bar X X + y_X \phi \bar X X \ .
\end{equation}
The SM Higgs doublet cannot couple to the $SU(2)_L$ singlet $X$ via renormalizable operators.

Loops of $X$ induce effective couplings between $\phi$ and gluons, photons and v-gluons. In the notation of \cite{Carmi:2012yp}:
\begin{eqnarray}
\nonumber 
\mathcal{L}_\mathrm{eff} &=& \tilde c_s \frac{\alpha_s}{12 \pi v} \phi G_{\mu \nu} G^{\mu \nu}
+
\tilde c_\gamma \frac{\alpha}{\pi v} \phi A_{\mu \nu} A^{\mu \nu}
\\
&& + 
\tilde c_v \frac{\alpha_v}{12 \pi v} \phi G'_{\mu \nu} G'^{\mu \nu}
\label{e.Leff}
\end{eqnarray}
where $G_{\mu \nu}$, $A_{\mu \nu}$ and $G'_{\mu \nu}$ are the gluon, photon, and v-gluon field strengths respectively. This effective field theory (EFT) description is valid as long as $m_\phi \ll 4 m_X$. The coefficients can be obtained by a simple rescaling of SM results:
\begin{equation}
\tilde c_g =  \tilde c_v = v \frac{y_X}{m_X} N_v \ \ \ , \ \ \tilde c_\gamma = \frac{ \tilde c_g }{2} Q_X^2 \ ,
\end{equation}
where $N_v = 3$ is the additional $SU(3)_v$ color factor.

The production cross section and decay widths of $\phi$ can now be straightforwardly calculated using \eref{Leff}. At 13 TeV, the production cross section for $m_\phi = 750 \gev$ in the EFT approximation is 
\begin{equation}
\sigma_{13 \tev} \  \approx  \ (100 \mathrm{fb}) \  y_X^2 \  \left( \frac{1 \tev}{m_X} \right)^2
\end{equation}
The decay widths to gluons, photons, and v-gluons (ignoring hadronic effects, like mass of the glueballs) is approximately
\begin{eqnarray}
\Gamma_{gg} &\approx& (13 \mev)   \  y_X^2 \  \left( \frac{1 \tev}{m_X} \right)^2
\\
\Gamma_{\gamma \gamma} &\approx& (0.4 \mev)   \  y_X^2 \  \left( \frac{1 \tev}{m_X} \right)^2 \ Q_X^4
\\
\Gamma_{g'g'} &\approx& \Gamma_{gg} \ \left(\frac{\alpha_v}{\alpha_s}\right)^2
\end{eqnarray}
where we have included the color factor $N_v =  3$, and in the last line the couplings are evaluated at scale $\mu = m_\phi$. Note that decay to v-gluons will not be kinematically available if the glueball mass is heavier than about 380 GeV.

\begin{figure}
\begin{center}
\includegraphics[width=0.4\textwidth]{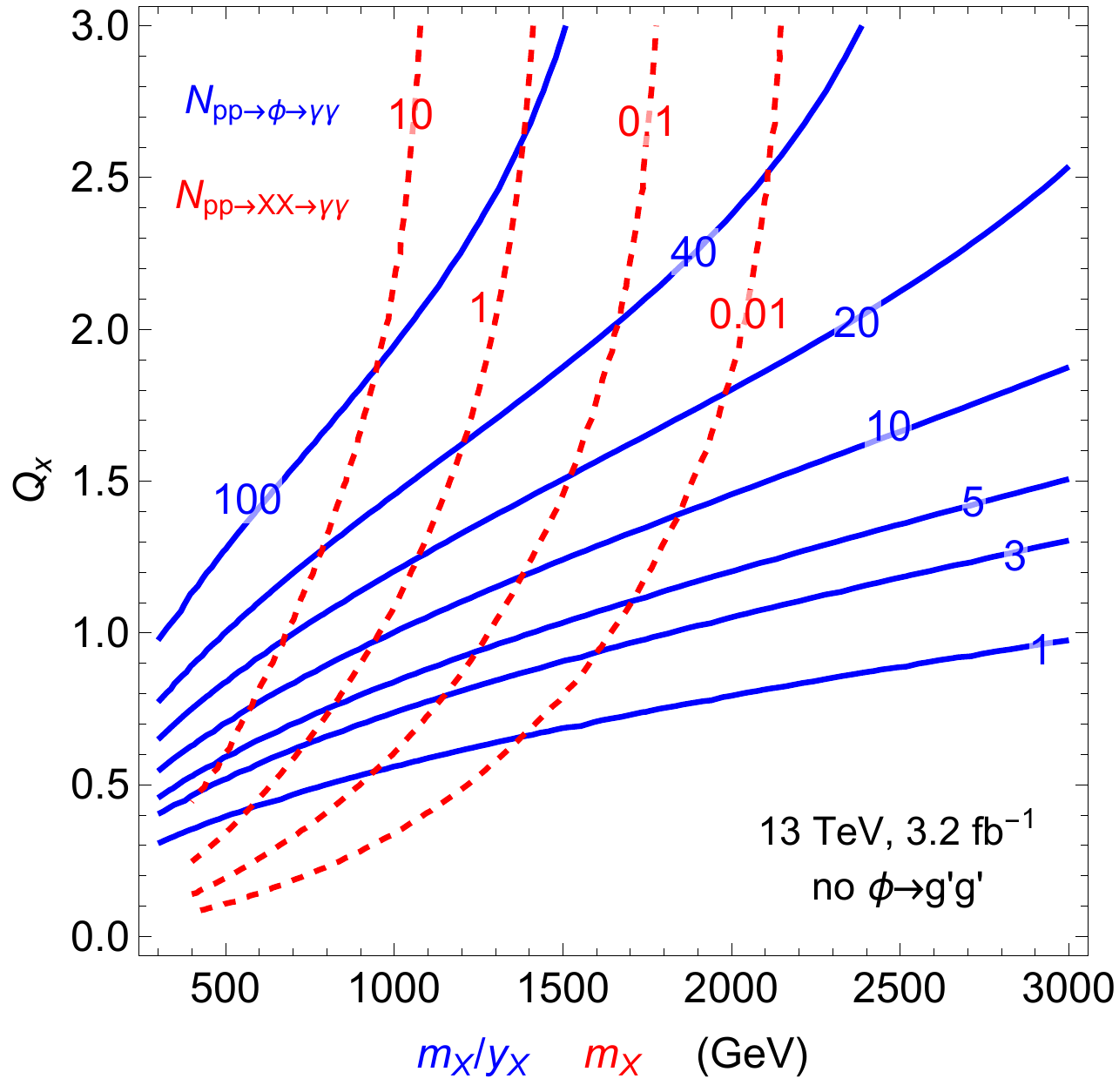} \vspace{5mm}
\\ 
\includegraphics[width=0.4\textwidth]{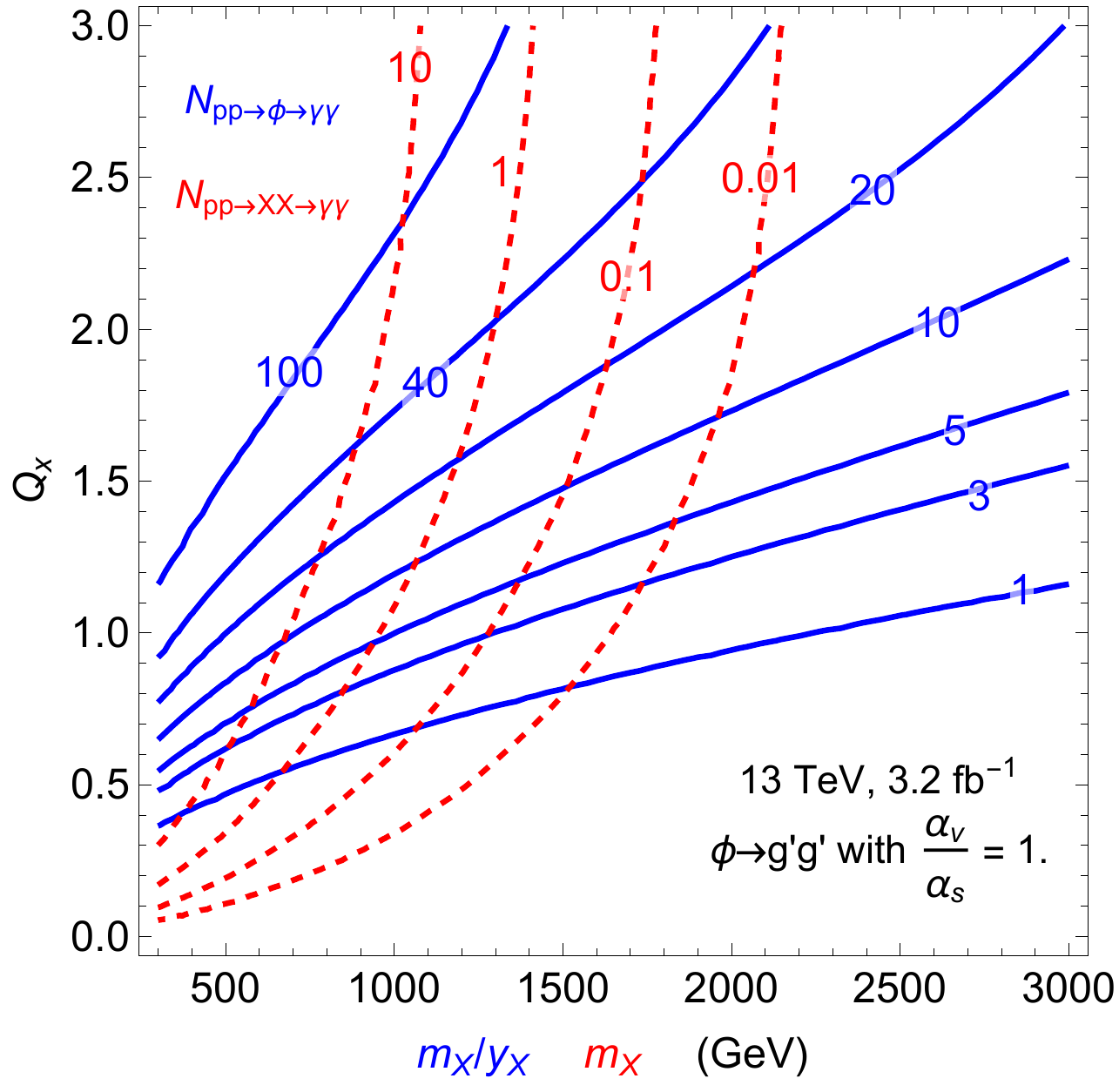}
\end{center}
\caption{
Blue contours: 
Number of diphoton events generated by our model at the 13 TeV LHC with 3.2 $\ifb$ of luminosity, as a function of $Q_X$ and $m_X/y_X$. The ATLAS excess favors $\sim 20 - 40$ events.
Red dashed contours: number of diphoton events with $m_{\gamma \gamma} \approx 2 m_X$ created from $\bar X X$ pair production, assuming the bound state annihilates dominantly in the spin singlet state, as a function of $Q_X$ and $m_X$ (note different definition of horizontal axis). 
Top: assuming $\phi$ cannot decay to v-gluons due to a large glueball mass. 
Estimate of $N_{p p \to X X \to \gamma \gamma}$ assumes $\alpha_v/\alpha_s = 1$, which corresponds to glueball masses in the 10 - 100 GeV range. Bottom: allowing decay to v-gluons with $\alpha_v/\alpha_s = 1$.
}
\label{f.Ngaga}
\end{figure}

\fref{Ngaga} shows the number of diphoton events from $\phi$ decay in the $(m_X/y_X, Q_X)$ plane at the 13 TeV LHC with $3.2 \ifb$ of luminosity (blue contours). In the top plot we assume $\phi$ decay to v-glueballs is kinematically forbidden. In the bottom plot we allow $\phi \to g' g'$ with $\alpha_v/\alpha = 1$, which corresponds to $\mathcal{O}(10 - 100 \gev)$ glueballs, see \fref{alphavoveralphas}.
The ATLAS excess \cite{atlas750} favors $\sim 20 - 40$ events, which can be achieved for $m_X/y_X \sim \tev$ and $Q_X \sim 1.5$. 

The detailed prediction for the number of diphoton events is subject to hadronization uncertainties in $\phi$ decay to both gluons and v-gluons, but this schematic estimate shows that the Quirk-Scalar model can easily generate the excess of observed diphoton events. Note that the total decay width of $\phi$ in this model is always very small. Therefore, if a significant decay width for $\phi$ was confirmed it would disfavor this minimal setup.

\textbf{\emph{Additional Signals of the Quirk-Scalar Model --- }}
In \fref{Ngaga} we also show, now in the $(m_X, Q_X)$ plane for $\alpha_v/\alpha_s = 1$, the number of diphoton events with $m_{\gamma \gamma} \approx 2 m_X$ expected in current run 2 data from quirk pair production and annihilation (dotted red contours), assuming annihilation is dominated by the spin singlet quirk state. For $m_X \gtrsim \tev$ there is currently little appreciable photon signal from quirk annihilation, but this could be detectable with $300 \ifb$ or even  $20 \ifb$ of data. 

Current 13 TeV dijet resonance searches \cite{ATLAS:2015nsi, Khachatryan:2015dcf} are still compatible with $m_X > 1 \tev$ but may be sensitive to quirk annihilation into SM gluons for $m_X \lesssim 1.5 \tev$ with $300 \ifb$ of data, depending on $\alpha_v/\alpha_s$.

We can again use dilepton searches to place a constraint on $R_{3/1}$, the fraction of quirks that annihilates as the spin triplet state. Due to the higher quirk mass, the most important constraint is derived from the 13 TeV bounds \cite{atlas8tevdilepton, CMS:2015nhc}. For $\alpha_v/\alpha_s$ in the $0.7 - 2$ range, the current run 2 data has no sensitivity to quirk annihilation for $m_X \gtrsim 1.2 \tev$, even for $R_{3/2} = 1$ and $Q_X \sim \mathcal{O}(1)$. With $\sim 300\ifb$ of run 2 data, dilepton resonance searches would only be sensitive to order unity $R_{3/1}$ for $m_X \lesssim 1.5 \tev$. Therefore, while a future dilepton signal is certainly possible in this Quirk-Scalar model, it is also easily possible to generate the 750 GeV diphoton excess without being excluded by dilepton bounds in the foreseeable future.

We now discuss the glueball signatures of the Quirk Scalar Model.  \fref{glueball} makes clear that a wide range of glueball phenomenology is possible. Very conservatively, these glueballs are cosmologically safe for $m_0 \gtrsim 50 \gev$ if $m_X \sim \tev$ (though these bounds may be somewhat relaxed depending on the relative abundances of glueballs in the early universe for $T < \Lambda_v$). Importantly, this cosmological bound implies that the $0^{++}$ decay length cannot be significantly longer than about a km. 

Glueballs can be produced both through quirk direct production, and in a large fraction of $\phi$ decays. If the Quirk-Scalar model explains the diphoton excess, then $\mathcal{O}(100)$ $\phi$ production events have already occurred at run 1 and run 2.\footnote{For $m_X \gtrsim 1 \tev$ (which is implied by the presumed lack of a $\gtrsim 2 \tev$ diphoton resonance in the 13 TeV data, see \fref{Ngaga}), the cross section for $\phi$ production is greater than for $\bar X X$ production as long as $y_X \gtrsim 0.5$.} If a $\mathcal{O}(1)$ fraction of these $\phi$'s decays to v-glueballs, then the fraction of events in which these v-glueballs decay with measurable displacement in the detector cannot be larger than $\mathcal{O}(1-10\%)$ \cite{CMS:2014wda}. This could be easily accommodated by long decay lengths, in which case additional luminosity will eventually reveal these glueballs in displaced dijet searches. A wildcard in these predictions are the shorter-lived glueballs like $1^{--}$ which can yield dilepton final states. Their production fraction in v-hadronization is unknown, but their decay, displaced or prompt, could be another spectacular signature. 

We also point out that the glueball lifetimes we have calculated could be reduced if $\phi$ were allowed to mix with the Higgs via the Higgs portal operator $\kappa |H|^2 \phi^2$, allowing additional mixing-suppressed operators for glueball decay \cite{Juknevich:2009gg}. However, the size of $\kappa$ is severely constrained to forbid large decays of $\phi$ to SM particles, so we do not consider this possibility here.

\textbf{CONCLUSIONS --- }
In this paper, we make the general point that  quirks, i.e. additional SM-charged fermions\footnote{Scalars are of course also a possibility.} interacting under their own confining gauge group, can be responsible for the diphoton excess reported by ATLAS \cite{atlas750} and CMS \cite{CMS:2015dxe}. In the first pure-quirk model we examined only the quirks are responsible, which is under tension from several other experimental constraints. In the second  Quirk-Scalar model, the quirks serve to ``hide'' the heavy SM-charged fermions required in the simple $\phi \bar X X$ toy model for the diphoton excess. This ameliorates the cosmological problems on heavy stable colored fermions, and gives rise to a pure glue Hidden Valley with quirky phenomenology. The notable LHC signals of the Quirk-Scalar model are a narrow width for the scalar $\phi$ and quirky pair production of a $\bar X X$ bound state which annihilates to v-gluons, SM gluons and photons, as well as possibly SM fermions. In addition to the 750 GeV diphoton resonance, this model may generate additional diphoton, dijet or dilepton resonances around $\gtrsim 2 \tev$ from quirk annihilation, as well as v-glueball production from  either $\phi$ decay or quirk annihilation. The v-glueballs can give rise to displaced vertices, lepton pairs, and photon pairs at the LHC run 2, which can be picked up by a variety of searches. Overall, if the Quirk-Scalar toy model is responsible for the 750 GeV diphoton excess, we would expect it to show up in one of these additional channels with 20 or $300 \ifb$ of LHC run 2 data.

\textbf{\emph{Acknowledgements ---}} We thank 
Kaustubh Agashe, 
Zacharia Chacko, 
Sungwoo Hong, 
Jose Juknevich,
Shmuel Nussinov,
Prashant Saraswat, and
Raman Sundrum
for useful discussion. 
D.C., and C.V. are supported by National Science Foundation grant No. PHY-1315155 and the Maryland Center for Fundamental Physics.

\bibliography{quirkydiphoton}

\end{document}